# The Accelerating Universe and the Second Law

M. Paul Gough

Space Science Centre, University of Sussex, Brighton, BN1 9QT, United Kingdom
E-mail: m.p.gough@sussex.ac.uk

**Abstract:** The main sources of information energy in the universe are shown to be stellar heated gas and dust and black holes. Information energy has properties similar to dark energy with a significant energy density that has remained nearly constant for at least the last half of cosmic time, corresponding to an equation of state parameter, $w_i$~-1, the value unique to dark energy. Changes in universe information content during star formation require an accelerating universe expansion in order to ensure $\Delta I \geq 0$. The size of this required acceleration is in good agreement with the observed extra doubling in universe size due to dark energy. Any information energy contribution to dark energy is determined by the extent of star formation, possibly answering the 'cosmic coincidence' question – "Why now?"

## 1. Introduction

The information present in all physical systems is directly bound up with the fundamental physics of nature. The reader is reminded of computer scientist Rolf Landauer's maxim "information is physical", and astrophysicist John Wheeler's slogan "it from bit".

Information entropy and thermodynamic entropy are identical when the same degrees of freedom are considered. A bit of information is therefore equivalent to $\Delta S = k_B \ln 2$ of thermodynamic entropy and the 2$^{nd}$ law of thermodynamics, $\Delta S \geq 0$, then also states that total information, $I$, never decreases, $\Delta I \geq 0$. This leads directly to Landauer's principle that any erasure of information within a system must cause that system to dissipate $\Delta S\ T = k_B\ T \ln 2$ of heat per erased bit into the environment surrounding that system, increasing the entropy of that environment so that overall $\Delta I \geq 0$, [1-14]. Landauer's principle is in effect only a restatement of the 2$^{nd}$ law, but nevertheless provides us with a useful approach to problems. For example, Landauer's principle has allowed us to finally reconcile Maxwell's Demon with the 2$^{nd}$ law [3, 9, 10, 14].

This paper considers information in the universe and extends previous work [15]. The energy contribution of information is considered to be the amount of heat dissipated via Landauer's principle if that information were to be erased. This information energy equivalence is similar to the standard cosmology practice of using $mc^2$ to represent the energy contribution of matter even though little mass has been converted to energy via nuclear fusion to date. For any system the information energy is then given by $N\, k_B\, T\, ln2$, where $N$ is the number of bits of information contained in the system, and $T$ the absolute system temperature.

## 2. Information Content and Information Energy Contributions to the Universe

The main contributors to universe information content are listed in Table 1 amalgamating two recent reviews [16][17], and adding typical temperatures to estimate information energy contributions.

|  | Information, $N$ Bits | Temperature, $T$ $^{\circ}$K | Information energy $N\, k_B\, T\, ln2$, Joules |
|---|---|---|---|
| $10^{22}$ stars | $10^{79} - 10^{81}$ | ~$10^7$ | $10^{63} - 10^{65}$ |
| Stellar heated gas and dust | ~$10^{86}$ | ~$10^6$ | ~$10^{69}$ |
| Relic gravitons | $10^{86} - 6 \times 10^{87}$ | ~1? | $10^{63} - 6 \times 10^{64}$ |
| Dark matter | ~$2 \times 10^{88}$ | <$10^2$ ? | <$10^{67}$ |
| CMB photons | $10^{88} - 2 \times 10^{89}$ | 2.7 | $3 \times 10^{65} - 6 \times 10^{66}$ |
| Relic neutrinos | $10^{88} - 5 \times 10^{89}$ | 2 | $2 \times 10^{65} - 10^{67}$ |
| Stellar black holes | $10^{97} - 6 \times 10^{97}$ | ~$10^{-7}$ | $10^{67} - 6 \times 10^{67}$ |
| Super massive black holes | $10^{102} - 3 \times 10^{104}$ | ~$10^{-14}$ | $10^{65} - 3 \times 10^{67}$ |
| Holographic upper bound | ~$10^{123}$ | - | - |

**Table 1.** Information content, temperature, and information energy contributions.

Single super massive black holes (~$10^7$ solar masses) that occur at the centre of most galaxies and other stellar sized black holes dominate the information bit content contributions to the universe (left-hand column of Table 1). Classically a black hole has a temperature at absolute zero emitting no radiation, but, in quantum theory, Hawking radiation is emitted with a perfect Planck spectrum that corresponds to an extremely cold object. For example, a stellar sized black hole typically has a temperature of only ~$10^{-7}$ of a degree above absolute zero. The larger super massive black holes at galactic centres have even lower temperatures ~$10^{-14}$ degrees. Table 1 shows that, despite their massive information contents, black holes do not over dominate the information energy contributions (right-hand column of Table 1) as much on account of their exceedingly low temperatures.

Only relatively weak information energy contributions are supplied by the low temperature relics of the big bang. Cosmic Microwave Background (CMB) decoupled from matter at the universe age of $3 \times 10^5$ years and maintained effectively constant information since then, with the CMB wavelength expanding in proportion to universe size, cooling to the present temperature of 2.7K.

Relic neutrinos and gravitons decoupled at a very early time and have temperatures today ~2K and ~0.6K, respectively. We can also expect the dark matter contribution to be low if we assume a cold dark matter model.

Table 1 shows the main information energy contributors to be primarily stellar heated gas and dust at a mean gas temperature of $10^6$K [18] and, to a lesser extent, black holes. Both major sources of information energy depend on the overall extent of star formation as this determines the amount of stellar heating of gas and dust, the formation of black holes from stellar collapse and associated heating of matter falling towards black holes. While there is considerable uncertainty in information content values used in Table 1, we can conclude that the information energy from major contributors will have varied over time essentially proportional to star formation and hence to the fraction of all baryons that are in stars.

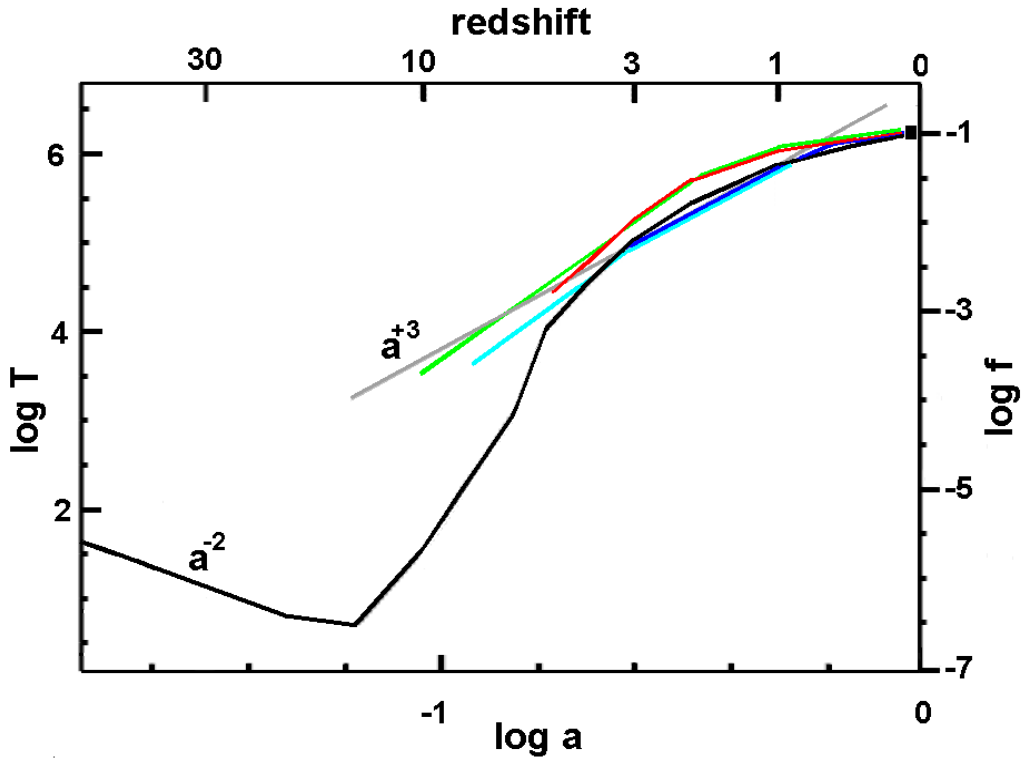

**Figure 1.** Combining measurements and models for the average baryon temperature, $T$, with the fraction of all baryons in stars, $f$, as a function of universe scale size, $a$.
Legend: Black: Mean gas temperature using GADGET model, Nagamine et al 2004[18]; Green: Stellar fraction from a survey at high $z$, Hu & Cowie, 2006[19]; Dark blue: Stellar density measurements, Hopkins & Beacom, 2006[20], Dickinson et al, 2003[21], Brinchmann & Ellis, 2000[22]; Red: Stellar mass density from infrared galaxies, Le Borgne et al, 2009[23]; Light blue: Modelling, Rasera & Teyssier, 2006[24]; Black square: Galactic fraction today, Cole et al, 2001[25]; Grey: The specific function $a^{+3}$ is shown for comparison.

As the stellar core temperatures, $\geq 10^7$K, are significantly higher than most non-stellar baryon matter, the average baryon temperature of the universe at any one time is proportional to the fraction of all baryons that are in stars [15]. This enables us to plot in fig.1 measured values and model predictions [18-25] for both average baryon temperature and baryon stellar fraction (from stellar densities) together on a single figure against universe scale size, $a$. See the caption of fig.1 for a list of data sources.

## 3. Information Equation of State Parameter

The time history variation of any contribution to the universe energy budget is described by an energy density varying as $a^{-3(1+w)}$ where $a$ is the universe scale size, and $w$ is the equation of state parameter. ($w_m = 0$ for matter, $w_{em} = +\frac{1}{3}$ for electromagnetic radiation, and $w_{de} = -1$ for dark energy contributions with dependances: $a^{-3}$, $a^{-4}$, and $a^0$, respectively)

We can assume that the amount of stellar heating of gas and dust is proportional to the average baryon temperature, $T$. Similarly the number of black holes generated by stellar collapse is proportional to the number of available stars given by the level of star formation, $f$. In Fig. 2 we plot the information equation of state parameter, $w_i$, that directly follows from the time history of $T$ and $f$ in Fig.1. Initially the information equation of state parameter, $w_i$, was positive, $w_i \sim +0.67$, as the universe cooled adiabatically, $T$ varying as $a^{-2}$ [18]. Later, as stars formed there was a long period, redshifts $8 > z > 0.8$, over one half of cosmic time, when $w_i \sim -1$, the key property unique to dark energy. Both $f$ and $T$ increased with increasing star formation approximately as $\sim a^{+3}$ (see gray line fig.1) compensating for the $a^{-3}$ universe dilution through expansion to provide a near constant information energy density, $a^0$.

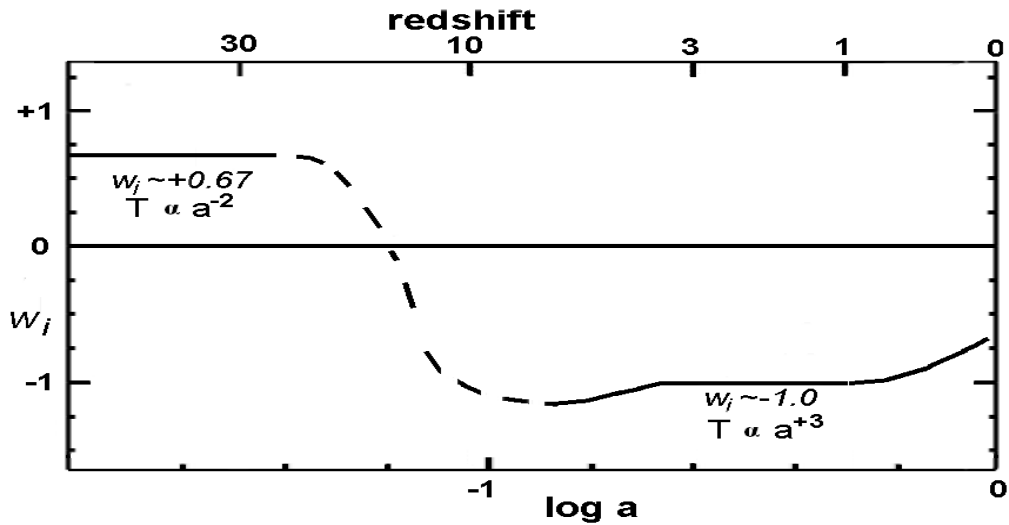

**Figure 2.** Information Equation of State Parameter, $w_i$, as a function of universe scale size, $a$.

This work improves on previous estimates [15][26] for the information equation of state parameter by having considered the wider range of possible contributors listed in Table 1 and by using the wider range of datasets and models included on Fig.1. In particular, this work identifies stellar heated gas and dust and black holes as the main sources of information energy.

Any negative valued equation of state, and, in particular, a specific value $w_i$~-1 implies that information energy contributes to dark energy. The total information energy value ~$10^{69}$ Joules from Table 1 is significant, at least a few percent of the present dark energy value. Note that the values used in Table 1 were adapted directly from [16][17] while other estimates [27-29] place total information contents, ~$10^{90}$ bits, a few orders of magnitude higher, for non black hole contributions. Given the uncertainties in the information content values used above, and noting the unique value of $w_i$~-1, information energy could therefore be making a major contribution to dark energy.

Also note that the average $k_B\ T\ ln2$ energy per bit has been calculated previously for information in a simple cooling universe with the same mass density as our universe today but without star formation [15][26]. This bit energy has the identical value today, ~3 x $10^{-3}$eV, and identical equation, as the characteristic energy of a cosmological constant, one of the proposed explanations for dark energy. Previously this low characteristic energy was difficult to relate to particle physics, but may now be explained as the average bit equivalent energy.

## 4. Hypothetical Computer Simulation Analogy

The previous section showed that information energy associated with star formation can make a significant contribution to the dark energy that causes the universe expansion to accelerate. It is therefore interesting to consider how the change in universe information content when stars are formed conforms to the second law, $\Delta I \geq 0$.

Note that the 2$^{nd}$ law does not necessarily require the universe to expand [30]. Also today's universe is very many orders of magnitude below the information limit given by the holographic bound that corresponds to the universe being one gigantic black hole (see Table 1)[29][31]. Nevertheless, in our computer simulation analogy below we see that the baryon information content depends on the size of the universe and on the extent of star formation.

We can obtain a good approximation to information variation over time by considering the amount of information that a hypothetical super computer would require in order to make a full 3-D simulation that follows the complete physics of the movement and interactions of each baryon. Clearly such an actual super computer would itself be many universes in size, so we must limit ourselves to a Gedanken(thought) experiment. We concentrate on the significant relative changes in information, without recourse to absolute information values, by employing the simple device of just considering what information accuracy would be needed for data input to our super computer.

For a guaranteed full physics simulation we require that each spatial parameter of every baryon be registered with a minimum accuracy of the Planck length, $1.6 \times 10^{-35}$m. For example, simulating intergalactic baryons in the present universe, size $\sim 10^{27}$m, then requires an accuracy of 1 part in $6 \times 10^{61}$ ($\sim 2^{205}$) and hence 205 bits per spatial parameter. Table 2 lists the typical sizes and number of bits per spatial parameter required for a range of relevant object types within the universe. A majority of the intergalactic baryons reside in filaments in the intergalactic medium while, within galaxies, giant molecular clouds are the stellar nurseries in which protostellar nebula form to condense into stars.

|  | Typical size, m | Bits per spatial parameter |
|---|---|---|
| Universe | $10^{27}$ | 205 |
| Intergalactic filaments | $3 \times 10^{22}$ | 190 |
| Galaxies | $10^{21}$ | 185 |
| Giant molecular clouds | $10^{18}$ | 175 |
| Protostellar nebula | $10^{15}$ | 165 |
| Stars, e.g. sun | $10^{9}$ | 145 |

**Table 2.** Information per dimensional parameter required for Planck length resolution.

Our model is primarily concerned with information changes during the period of star formation so we can ignore the relics of the big bang listed in Table 1. CMB, relic gravitons and relic neutrinos all have essentially constant information as they have had little interaction with matter during this period. For example, the resolution required to describe CMB is naturally the CMB wavelength which increased in direct proportion the universe scale size, $a$, ensuring a constant information content. There is a very weak imprint of large groups of galaxies on the CMB temperature (the Sunyaev-Zel'dovich effect, $\delta T/T \sim 10^{-4}$, [32]) but this should provide a negligible contribution to the overall CMB information.

In the following simple model all intergalactic baryons are located in intergalactic filaments and the remaining galactic baryons are either located in giant molecular clouds or in stars. In fig.3 the fraction of all baryons that are in stars, $f$, is combined with the expanding scale size, $a$, of the universe.

In this model the 10% of baryons that are found in stars today require spatial parameters in our hypothetical computer simulation that fell in size by 30bits from the 175bits/parameter corresponding to giant molecular clouds to the 145bits/parameter of stars over this period following the variation of the stellar fraction, *f*, from fig.1. The remaining 90% of baryons, located in cosmic web filaments that scale in proportion to the universe, are described by spatial parameters that increased in size by one bit for every factor of two increase in universe size, *a*, up to their present value of 190bits.

In fig. 3 the overall average number of bits per baryon spatial parameter increases with increasing universe size to reach a maximum before the information reduction from increasing star formation causes a fall by 0.9 bits to the present value (dashed line top panel fig.3). This ~1bit loss is the result of changed contributions to the average baryon parameter over this period: a loss of ~3 bits (10% of 30 bits) for the 10% of baryons forming stars, against a gain of only ~2 bits for the 90% intergalactic baryons as the universe expanded by a factor of four over this period from around *z*~3 to the present.

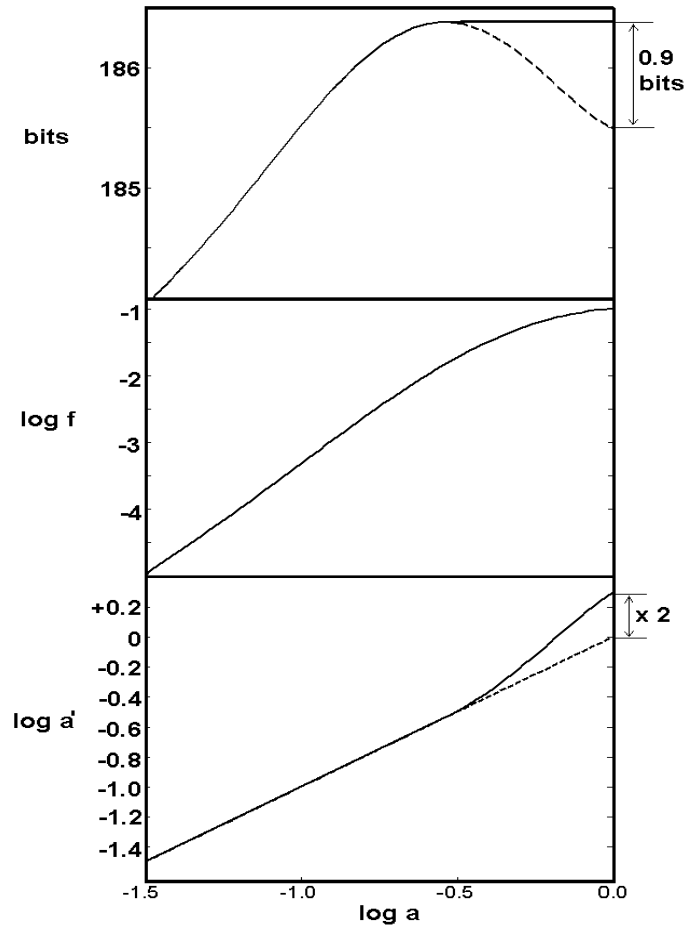

**Figure 3. Top panel**: Average number of bits per dimension per baryon for a full physics computer simulation. Dashed line: non-accelerated expansion. Full line: extra, accelerated expansion required for $\Delta I \geq 0$. **Middle panel**: Fraction, *f*, of all baryons that are in stars (from data in fig.1), **Lower panel**: Corrected universe scale size, *a'*, i.e. corrected to ensure $\Delta I \geq 0$, corresponding to the horizontal line in the top panel.

Our model has not included the vast reservoirs of information represented by the black holes that have formed over this period, see Table 1. However, this black hole information remains inside the black hole's event horizon and thus invisible to, and unavailable to, the universe as a whole. From the universe's point of view a black hole can be <u>fully</u> specified by just three parameters: mass; angular momentum; and charge. Effectively black hole information will have no significant effect on our calculation as it is equivalent to the very small fraction of stars that have collapsed to form black holes appearing to have average baryon spatial parameters of a fraction of a single bit instead of ~145bits.

Returning to fig 3, a fall in the quantity of information required for a full baryon physics simulation means there must be a corresponding loss of information in the real universe under simulation, contravening the 2$^{nd}$ law of thermodynamics, $\Delta I \geq 0$. The only way to remove the above fall of 0.9 bits, within the simple framework of our hypothetical computer simulation, is to assume that the universe has recently expanded at a faster rate to provide more available states in the intergalactic medium. The corrected scale size, *a'*, in fig 3 lower panel, is the minimum size that ensures zero information loss, horizontal line top panel fig 3, and compliance with the 2$^{nd}$ law. We see from fig 3 that the required accelerated expansion would lead to approximately an extra factor of two increase of scale size by today.

This factor of two agrees well with the actual accelerating expansion. The dark energy proposed to explain accelerated expansion has been determined to be presently three times the energy density of all matter. Hence the Hubble parameter, *H*, is approximately a factor of two greater today than it would have been if the universe had expanded without dark energy. ($3H^2 \approx 4 \times 8\pi G\rho$ with dark energy instead of $3H^2 \approx 8\pi G\rho$ without dark energy, where *G* is the Gravitational constant, and $\rho$ the universe mass density including dark matter [33].) Thus the actual accelerated expansion fits the amount required to adhere to the 2$^{nd}$ law. We conclude that the observed accelerating expansion could be an essential requirement for the universe to comply with the 2$^{nd}$ law during star formation. It is important to reiterate here that this is solely a mechanism to generate the extra states required to ensure $\Delta I \geq 0$. The universe total information content is far below any limit and, in any case, the 2$^{nd}$ law does not necessarily demand an expanding universe [30].

The above model is simple and robust – it is not overly sensitive to the parameter values chosen. For example, if the intergalactic baryons were considered to be free within the universe rather than primarily tied to filaments (i.e. 205 bits instead of 190 bits today), or if we used a much coarser resolution, say of the order of atomic nuclei instead of the Planck length (i.e. ~10$^{-15}$m instead of 1.6 x 10$^{-35}$m), then only the absolute bit values of fig.3 top panel would change. The 0.9 bit difference and the required extra expansion factor of two would remain unchanged. Critical factors that affect our result are the fraction, *f*, in recent times, at redshifts 5>*z*>0, where there is clearly good agreement between the various measurements and models in fig.1, and the amount of information lost during star formation. This information loss must be ~30bits, as it is set by the nine orders of magnitude difference in sizes between the start and end points of star formation: typical giant molecular clouds and typical stars. We have considered all baryons in the intergalactic space, in molecular clouds, and in stars. With the exception of black holes, any other baryon populations, within galaxies but not involved in star formation, can be assumed to have near constant information over this period and hence not significantly affect our results.

## 5. Discussion and Summary

In sections 2 & 3 we saw that black holes may contribute significant dark information energy but in section 4 we assume their information content has no impact on how total universe information adheres to the $2^{nd}$ law. This apparent contradiction arises from the unique properties of black holes. For example, a super massive black hole is known to exist at the centre of our galaxy from the stars orbiting it. It will have an enormous internal information content (see Table 1) set according to the holographic principle by its surface area, set in turn by its mass. Despite its high internal information content, and despite the significant influence it exhibits on those orbiting stars, the only knowledge the universe has of that black hole is completely described in just three parameters. Two black holes with the same mass, spin, and charge would be absolutely <u>identical</u> as far as the universe is concerned. Thus the single $\sim 10^7$ solar mass super massive black hole at the galactic centre provides insignificant information externally compared to the information of the galaxy's $\sim 10^{11}$ stars, gas and dust. Solar mass sized black holes have lifetimes $\sim 10^{67}$ years, orders of magnitude greater than the present age of the universe, eventually evaporating via Hawking radiation. In this way their information energy may count over time but, at any one time, at least at the present time, stellar sized black holes exhibit negligible information content to the rest of the universe. How, and whether, all of the massive internal information content returns to the universe at the end of a black hole's lifetime, the so called 'black hole information paradox', is therefore not relevant to our discussion here.

We have shown that the main contributions to the overall universe information energy, from stellar heated gas and dust and from black holes, vary with stellar formation as $\sim a^{+3}$. This results in an information equation of state parameter, $w_i \sim -1$, a value specific to dark energy, that has lasted at least the last half of cosmic time. Considerations of baryon information content during star formation are consistent with the dark energy related accelerating expansion providing the observed extra factor of two in size. We have found information energy has a similar effect on the universe to dark energy, exhibits similar properties, and requires a similar response from the universe in a self-consistent manner, in accord with the $2^{nd}$ law, both via Landauer's principle and the overall requirement, $\Delta I \geq 0$.

A complementary understanding is provided by applying Landauer's principle to the separate viewpoints of galaxies and the universe. Applying Landauer's principle to galaxies, this information 'erasure' during star formation within a galaxy should result in $k_B \, T \, ln2$ of heat per 'erased' bit being exported to the environment surrounding the galaxy. That export should heat up that environment and effectively increase the intergalactic medium information to compensate, resulting in zero overall information loss. On the other hand, from the universe's point of view, there is, by definition, no external environment to which the universe can export the $k_B \, T \, ln2$ heat per bit 'erased' inside it by star formation. Therefore the universe has to expand further to generate more states in the intergalactic medium within and ensure no loss of information. Clearly the extra states generated by the galaxies in their local intergalactic medium, and the extra states produced by the accelerating expansion of the universe also in the intergalactic medium, are one and the same, just seen from different viewpoints.

Information energy considerations presented in the above sections might therefore directly tie the recent accelerated expansion to the period of increasing star formation and hence provide an answer to the 'cosmic coincidence' question: "Why now? Why is the acceleration occurring in our epoch?"

This work may also provide a response to the question that must follow: "What next? Will the universe continue to accelerate forever?" The fraction, $f$, of baryons in stars is clearly reaching a plateau value just above 10% (see fig.1) and models of the average baryon temperature [18] show that the temperature will plateau to $a \sim 100$ before eventually falling adiabatically again as $a^{-2}$. So, in future, we expect any information energy related accelerating expansion to cease, returning towards a more constant expansion lasting perhaps until the universe has grown to a hundred times the present size. Already the start of a plateau in the value of $f$, evident in fig.1, should have caused some reduction of acceleration by now, an effect that should soon be detected as the sensitivity of dark energy measurements continues to improve.

To summarise we find that:

- The accelerating expansion can be tied to the 2$^{nd}$ law, both via $\Delta I \geq 0$, and via Landauer's principle from information content and information energy considerations respectively.

- The main contributors to the universe information energy are stellar heated gas and dust and black holes. Both contributions depend on the extent of star formation.

- Information energy could contribute significantly to dark energy as it has similar properties: a near constant energy density, $w_i \sim -1$; and a high value for a over the last half of cosmic time.

- Information content considerations during star formation are consistent with an accelerating universe that has lead to an extra factor of two in size today – similar to that observed.

- Star formation drives any dark information energy contribution to accelerating expansion, possibly providing an answer to the 'cosmic coincidence' question: "Why now?"

- As the fraction of baryons in stars is now increasing less steeply we predict higher resolution measurements will show reduced dark energy acceleration in the most recent time period.

**References**

1. Landauer, R. Irreversibility and Heat Generation in the Computing Process. *IBM J. Res. Dev.* **1961**, *3*, 183-191.
2. Bennett, C.H. Logical Reversibility of Computation. *IBM J. Res. Dev.* **1973**, *17*, 525-532.
3. Bennett, C.H. The Thermodynamics of Computation – a Review. *Int. J. Theor. Phys.* **1982**, *21*, 905-940.
4. Bennett, C.H. Notes on the History of Reversible Computation. *IBM J. Res. Dev.* **1988**, *32,* 16-23.
5. Landauer, R. Dissipation and noise immunity in computation and communication. *Nature* **1988**, *335*, 779-784.
6. Landauer, R. Computation: A Fundamental Physical View. *Phys. Scr.* **1987**, *35*, 88-95.
7. Bennett, C.H. Information physics in cartoons. *Superlatt. Microstruct.* **1998**, *23*, 367-372.
8. Feynman, R. P. *Lectures on Computation*; Penguin Books, **1999**; pp. 137-184.
9. Bennett, C.H. Notes on Landauer's Principle, Reversible Computation, and Maxwell's Demon. *Stud. Hist. Phil. Mod. Phys.* **2003**, *34*, 501-510.
10. Plenio, M.B.; Vitelli, V. The physics of forgetting: Landauer's erasure principle and information theory. *Contemp. Phys.* **2001**, *42,* 25-60.
11. Ladyman, J.; Presnell, S.; Short, A.J.; Groisman, B. The connection between logical and thermodynamic irreversibility. *Stud. Hist. Phil. Mod. Phys.* **2007**, *38*, 58-79.
12. Piechocinska, B. Information erasure. *Phys. Rev. A* **2000**, *61*, 062314:1-062314:9.
13. Daffertshofer, A.; Plastino, A.R. Landauer's principle and the conservation of information. *Phys.Lett. A* **2005**, *342*, 213-216.
14. Leff, H. S.; Rex, A. F., (Eds.) Maxwell's Demon 2: Entropy, classical and quantum information, computing. **2003**, CRC Press.(and papers therein).
15. Gough, M.P. Information Equation of State, *Entropy* **2008**, *10*, 150-159.
16. Frampton, P.H.; Hsu, S.D.H; Reeb D.; Kephart, T.W. What is the entropy of the universe?, **2009**, 5p., arXiv:0801.1847v3.

17. Egan, C.A.; Lineweaver, C.H. A larger estimate of the entropy of the universe, **2009**, 11p, arXiv:0909.3983v1
18. Nagamine, K; Loeb, A. Future evolution of the intergalactic medium in a universe dominated by a cosmological constant, *New Astronomy* **2004**, 9, 573-583.
19. Hu, E.M.; Cowie, L.L. High-redshift galaxy populations. *Nature* **2006**, *440*, 1145-1150.
20. Hopkins, A.M., Beacom, J.K., On the normalisation of the cosmic star formation history. *Astrophysical Journal* **2006**, 651, 142-154.
21. Dickinson, M.; Papovich,C.; Ferguson, H.C.; Budavari, T., The evolution of the global stellar mass density at 0<z<3, *Astrophysical Journal* **2003**, 587, 25-40.
22. Brinchmann,J.; Ellis,R.S., The mass assembly and star formation characteristics of field galaxies of known morphology, *Astrophysical Journal* **2000**, 536, L77-L80.
23. Le Borgne, D.; Elbaz,D.; Ocvirk,P.; Pichon, C. Cosmic star formation history from non-parametric inversion of infrared galaxy counts, *Astronomy and Astrophysics* **2009**, manuscript no LeBorgne09
24. Rasera, Y.; Teyssier, R. The history of the baryon budget, *Astronomy and Astrophysics* **2006**, 445, 1-27
25. Cole, S.; Norberg, P.; Baugh, C.M.; Frenk, C.S.; Bland-Hawthorn, J.; Bridges, T.; Cannon, R.;Colless, M.; Collins, C.; Couch, W.; Cross, N.; Dalton, G.; De Propris, R.; Driver, S.P.;Efstathiou, G.; Ellis, R.S.; Glazebrook, K.; Jackson, C.; Lahav, O.; Lewis, I.; Lumsden, S.; Maddox, S.; Madgwick, D.; Peacock, J.A.; Peterson, B.A.; Sutherland, W.; Taylor, K. The 2dF galaxy redshift survey:near-infrared galaxy luminosity functions. *Mon. Not. R. Astron. Soc.* **2001**, *326*, 255-273.
26. Gough,M.P.; Carozzi, T.; Buckley, A.M. On the similarity of Information Energy to Dark Energy. arXiv E-Print, **2006**; 6 p., [astro-ph/0603084].
27. Lloyd, S. Ultimate physical limits to computation. *Nature* **2000**, *406*, 1047-1054.
28. Lloyd, S. *Programming the Universe: A Quantum Computer Scientist Takes On the Cosmos*; **2006**, Alfred A. Knopf Publisher.
29. Lloyd, S. Computational Capacity of the Universe. *Phys. Rev. Lett*. **2002**, *88*, 237901:1-237901:4.
30. Penrose, R. *The Road to Reality*; Jonathan Cape: London, **2004**; pp. 705-707.
31. Bekenstein, J.D. Black Holes and Entropy. *Phys. Rev. D* **1973**, *7*, 2333-2346.
32. Sunyaev, R.A. and Zeldovich,Y.A.B. Small-scale fluctuations of relic radiation, *Astrophysics and Space Science* **1970**, *7*, 3-19
33. Räsänen, S., The effect of structure formation on the expansion of the universe, *Int. J. Mod. Phys* **2008**, *D17*, 2543-2548.